# Rethinking massive multiplexing in whispering gallery mode biosensing

Ivan Saetchnikov[1], Elina Tcherniavskaia[2], Andreas Ostendorf[3], Anton Saetchnikov[3*]

[1]Radio Physics Department, Belarusian State University, Minsk, 220064, Belarus.
[2]Physics Department, Belarusian State University, Minsk, 220030, Belarus.
[3*]Applied Laser Technologies, Ruhr University Bochum, Bochum, 44801, Germany.

*Corresponding author(s). E-mail(s): anton.saetchnikov@rub.de

**Abstract**

Accurate, label-free quantification of multiple analytes in complex biological media remains a major challenge due to limited multiplexing, signal cross-correlations, and inconsistency across sensor samples and measurement runs. We introduce a multiplexed whispering-gallery-mode (WGM) biosensing framework that overcomes these barriers by jointly advancing photonic integration and data analytics. Our glass-chip platform enables massive, parallelized and flexible multiplexing of >10000 microresonators organized into up to 100 sensing channels, with universal and modular chip design and detection hardware, while maintaining loaded Q-factors of $10^6$. Our novel hybrid deep-learning framework BioCCF that integrates domain adaptation with cross-channel fusion enables harmonization of responses across sensing chips and extraction of nonlinear correlations in complex mixtures. Using a highly heterogeneous dataset comprising over 200 hours of sensing data acquired from nine chips with different channel configurations, biological replicates, and repeated regeneration cycles, we demonstrate recalibration-free identification of solution (99.3% accuracy) and quantification of immunoglobulin G components with relative prediction error of $10^{-4}$ under 5 min. The affordability and modularity of the platform enable distributed data acquisition and aggregation into shared repositories, providing a pathway toward continuously improving model generalization, cross-validation and a scalable, community-driven paradigm for biosensing.

# Introduction

Biosensing is increasingly demanded by personalised medicine and environmental monitoring. Real samples usually contain multiple co-occurring analytes, which require sensitive, selective and miniaturised multiplexed detectors. For example, simultaneous detection of various immunoglobulins G (IgG) helps in assessing immune status and disease progression. Optical biosensors offer core advantages over labeled assays such as ELISA, including rapid, real-time, and highly sensitive detection. Whispering gallery mode (WGM) microresonators are highlighted among them by strongly confined optical field and high quality (Q)-factors, enabling precise real-time sensing [1–4]. Perturbations near the cavity, such as refractive index changes, adsorption, or scattering, produce WGM spectral shifts, linewidth changes, or mode splitting [5–7]. Coupling to plasmonic particles, gain-assisted/lasing schemes, or exceptional-point operation further boost WGM performance [8–14]. Microreosonators have already been proven as versatile probes of physical, chemical, and biochemical parameters [9, 15–21] also as embedded into organisms for intracellular sensing, multiplexed cell tracking, and deep-tissue imaging [22–25].

A primary bottleneck in WGM biosensing is scalable multiplexing with dense microresonator integration on a compact, low-cost platform suitable for point-of-care and field diagnostics [26, 27]. Beyond resolving components in complex solutions, multiplexing is crucial for monitoring factors that modulate bioreceptor–analyte affinity such as pH and temperature alongside extensive controls to verify specificity, eliminate false positives, and confirm bioreceptor functionality [28]. These demands make high-degree multiplexing the core for robust, reproducible biosensing. Although microcavities have a small footprint, their massive interrogation while preserving high Q-factors and sensitivity remains a major challenge, ultimately dictated by the geometry of the cavities and the light coupling strategy.

Multiplexing with tapered-fiber bus is achieved by sequentially placing fiber-melted microspheres in optical contact [29, 30]. This requires nanoscale alignment and exceptional stability (ensured by gluing) [31, 32] and have limited multiplexing due to cavity signal overlapping [33]. Disks, toroids, and goblets on pillars with high Q-factor improve fabrication scalability, but still rely on external tapered fibers [34, 35] or complex wafer-integrated fibers [36, 37]. Thus, the complexity drastically grows with cavity count and limits multiplexing to few units [38]. Free-space coupling is alternative to fragile tapers for microlasers with performance limited by detector spectral resolution rather than Q-factor [38–40]. Recently high Q-factors have been achieved for free-space coupling of passive cavities at efficiencies below 0.1% [41], but multiplexing remains limited by sequential cavity scanning. Silicon-on-insulator (SOI) rings offer high refractive-index contrast, CMOS compatibility and scalable fabrication [42, 43], but Si restricts visible-range operation to more complex $Si_3N_4$ platforms [44]. Although on-chip waveguides simplify coupling, Q-factors rarely exceed $10^6$ in air and degrade in aqueous media [45]. Moreover, simultaneous readout demands complex feed waveguide networks often sequentially interrogated by scanning grating couplers [46]. In summary, the integration complexity of known multiplexing schemes linearly scales with resonator number, often operating only in a sequence that fundamentally constrains multiplexing.

A desired real-time and massively multiplexed sensor inherently generates high-dimensional data with complex temporal nonlinear correlations, making artificial intelligence and machine learning (AI/ML) analytics essential. Since our first demonstration of ML-enabled sensing using narrow-band WGM spectrum [47], the field has progressed toward leveraging high-dimensional multimode spectra of single cavities [48–51]. We employed high dimensionality emerging by monitoring hundreds of microresonators at fixed laser wavelength to study static and dynamic processes [52, 53]. Data-driven models learn extracted informative features and hidden correlations, delivering better sensitivity and broader dynamic range. However, they assume that new measurements match the training data feature space, which ties model to a specific cavity or their fixed assembly. Biosensing datasets span different sensor samples, alignments, and noise conditions, where minor variations among them demand complete retraining, making existing ML approaches unfeasible for field deployment [54]. Furthermore, the high sensitivity of the microresonator makes its high-Q spectra unique by definition.

We introduce a breakthrough in multiplexed biosensing by synthesising an affordable, modular and densely packaged WGM glass chip with novel hybrid DL-engine that we named Biological Cross-Channel Fusion (BioCCF) (Fig. 1). The WGM chip supports integration of thousands of cavities with Q$\sim 10^6$ across 100 channels for simultaneous probing. Domain adaptation (DA) ensures sensorgram consistency among different chips, whereas DL-engine joins convolutional–bidirectional LSTM classifier with transformer-encoder regressor enhanced by cross-attention fusion for fast and accurate prediction of analyte concentrations. BioCCF integrates segments of sensorgrams across the channels, provides robustness to missing or irregular segments and learns correlations among channels, supports scalability by adding new control or sensing channels. Our immunoassay experiments demonstrates prediction on IgG components from complex mixtures in less than 5 min, achieving relative prediction error of $10^{-4}$ for concentrations up to ∼10 $\mu$g/ml using nine chips over numerous sensing runs. The proposed framework eliminates the requirement for chip-specific recalibration, enabling distributed measurements and sensing data aggregation that approaches real-world feasibility of WGM biosensing.

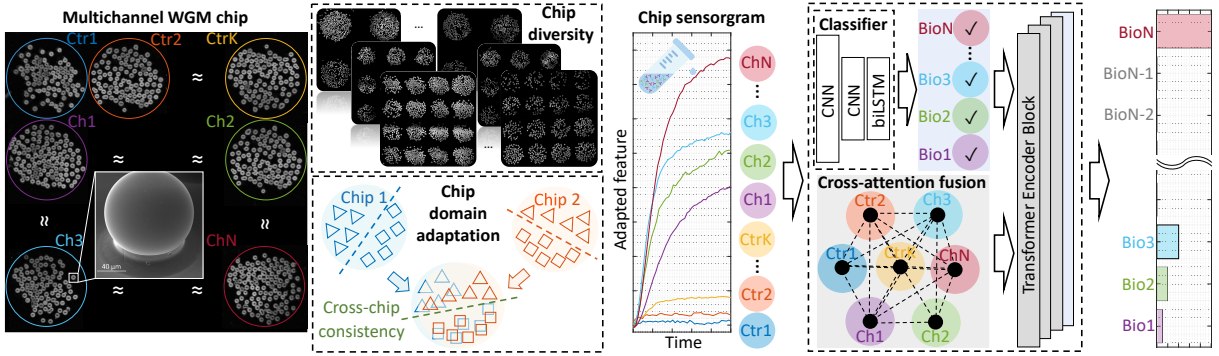

**Fig. 1**: **Concept of massively multiplexed biosensing with WGM microresonators.** Multichannel WGM chip containing thousands of permanently immobilized glass microspheres organized into N sensing (Ch1–ChN) and K control (Ctr1–CtrK) channels (N+K ≃ 100) and supports simultaneous signal collection. Chip-to-chip diversity is mitigated using domain adaptation, enabling consistent sensing responses and facilitating the construction of large aggregated datasets. A hybrid ML-framework BioCCF combining a CNN/biLSTM-based classifier with a cross-attention fusion learns from multichannel chip responses and enables accurate predictions of the individual substance concentrations (Bio1–BioN) in complex solutions.

# Results

## Massive WGM multiplexing

Detecting complex biochemical mixtures requires the number of sensing elements to match or exceed the number of components. We realize this by defining multiple sensing channels on a single glass chip, with microspheres deposited at predetermined locations using 3D-printed masks containing $n^2$ (n integer) pass-through cells. The microspheres are permanently immobilized by spin-coating a low-refractive-index polymer film (MY-133MC, MyPolymer). Exemplary chips comprising 4, 9, and 16 sensing channels populated with microspheres from a [106–125] $\mu$m batch are shown in Fig. 2a–c. For reliable WGM signals determination, cavity maps and cluster assignments are constructed for each chip (Supplementary Information, Sec. S1), allowing unambiguous identification of microresonators and their associated channels. Due to the random arrangement and batch-specific size distribution, the number of cavities varies both across channels and between chips. Statistical analysis of 24 chips shows that increasing the channel count from 4 to 16 reduces the median number of microresonators per channel from 235 to 100, while simultaneously decreasing inter-chip variance (Fig. 2d). Beyond 16 channels, the median cavity count from the batch [106–125] $\mu$m drops below 100, limiting sensing reliability (see Domain Adaptation).

The fixation layer is critical for signal stability, sensitivity, and sensor reusability being immersed in liquid. Insufficient thickness leads to unstable WGM excitation and poor reusability, whereas excessive thickness increases optical losses and degrades performance. In order to minimize coupling losses we controlled phase matching by the coupling angle, while the field overlap is fine-tuned by adjusting the prism–microresonator distance set by fixation layer. Numerical calculation, based on the model in [55], shows that a 100 $\mu$m glass microsphere exhibits coupling-limited Q-factor ($Q_{coup}$) at $10^3$ (680 nm) with zero gap. Increasing the gap to 350 nm raises $Q_{coup}$ to $10^6$, and the absorption-limited value for soda-lime glass ($Q_{abs} \sim 10^7$) is reached at 470 nm. To achieve a uniform sub-micrometer spacing, the fixation agent is diluted before spin-coating. The statistics of loaded Q-factor over distancing layer thickness is summarized in Fig. 2 e.

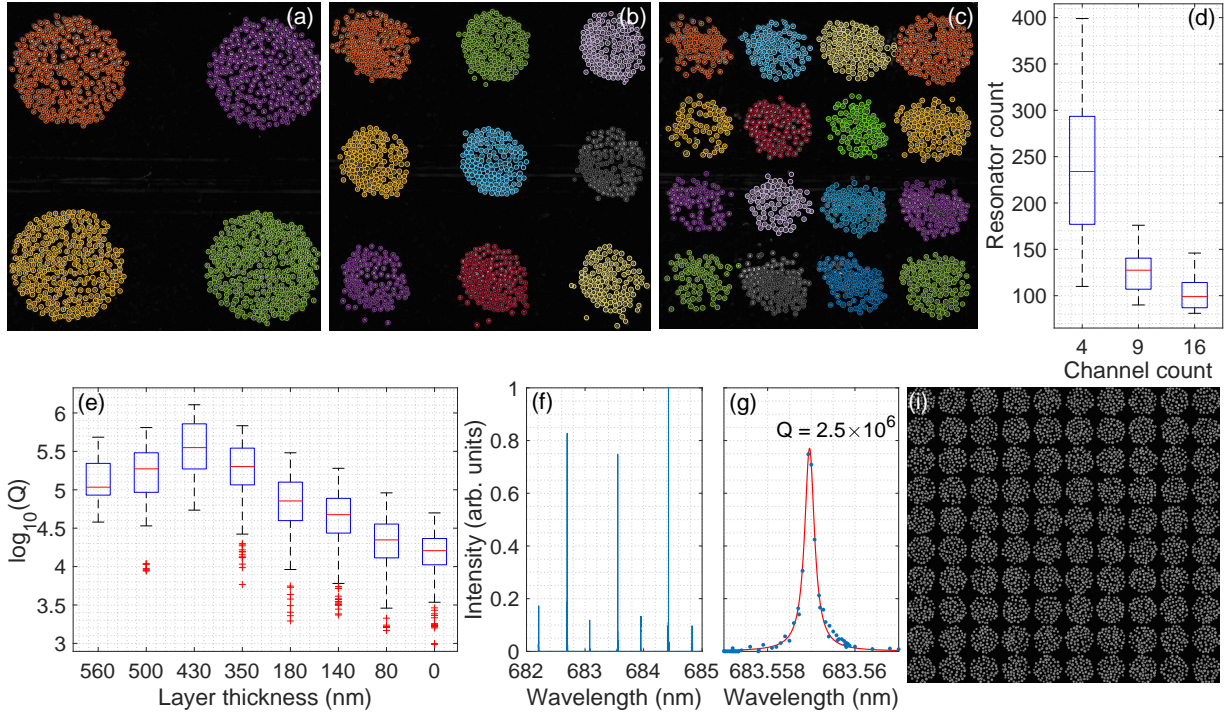

**Fig. 2**: **Characterization of the multichannel WGM sensor.** (a-c) Exemplary chips with 106:125 $\mu$m microspheres allocated in 4 (a), 9 (b) and 16 (c) channels. Each microcavity is channel-coded with color. (d) Statistics on microresonator filling depending on channel count. (e) Loaded Q-factor in water fabricated on the preprocessed substrate with distancing layer (80-560 nm) including non-processed substrate (0). (f) WGM spectrum over 3 nm range of a representative cavity with ≈430 nm distancing layer. (g) Resonance line along with estimations of the Q-factor achieved by fitting the experimental data (dots) with a Voigt profile (line). (i) Overview of high-throughput (100 channels) multiplexing with 63:75 $\mu$m microspheres.

Microresonators without a distancing layer (0 nm) exhibit a median Q of $1.6 \times 10^4$, with some cavities reaching $5 \times 10^4$ — well above the numerically predicted $10^3$. This discrepancy arises from variance of the cavity sizes and partial embedding in the fixation layer, which prevents full contact with the substrate for smaller spheres. Introducing a distancing layer notably increases the loaded Q, but the contrast of resonances may reduce. At ≈80 nm spacing, over 90% of resonators show clear WGMs whereas at ≈560 nm less than 5% remain active, while exhibiting Q-factors of $4 \times 10^4$-$5 \times 10^5$ with a median of $1 \times 10^5$. The layer of 430 nm provides optimal performance, balancing yield and Q-factor: resonances are observed for >60% of cavities, with median Q in the mid-$10^5$ range and maxima approaching $3 \times 10^6$. A representative resonance spectrum and a zoom in the resonance area of Q> $10^6$ in aqueous medium are shown in Fig. 2f–g. Reducing the microsphere sizes from [106:125] $\mu$m to [63:75] $\mu$m scales the platform to 100 sensing channels while preserving comparable Q-factors, with approximately 100 cavities per channel (Fig. 2i). This enables high-throughput operation with up to $10^4$ microresonators polled simultaneously.

## Domain adaptation

Despite massive and scalable integration, identical high-Q WGM chips cannot be produced, leading to strong cross-chip variability arising from different number of cavities per channel and inherently unique spectra of microresonators. Meaningful sensing therefore requires mapping heterogeneous measurements into a common feature space. Another aspect is the limited experimental throughput that constrains dataset volume and diversity, which must include both technical (same biological sample) and biological (biologically distinct samples) replicates. We address these challenges by converting spectrally-resolved measurements (acquired via laser sweep) into a unified intensity-based representation through normalization, spectral and temporal alignment. This transformation generates thousands of effective technical replicates per run, lowering data-volume constraints. However, spectral uniqueness of microcavities amplifies variability in the intensity domain, hindering direct pattern recognition. We propose domain adaptation to harmonize chip responses via projecting highly heterogeneous data into a shared latent space.

We performed domain adaptation (DA) independently for each single-wavelength dataset to study the wavelength independence of the channel response in the adapted domain. For each wavelength ($\approx$4800 with a step of 0.2 pm), the number of microresonators (features) is varied from 10 to 200 to assess the consistency of the adapted responses. We also added an ultimate case in which signals from all cavities at all wavelengths are merged as independent features ("Inf"), corresponding in fact to million features/cavities. We randomly split all datasets with different number of features into source and target subsets and repeated the procedure thousand times to ensure statistical robustness. The sensing response is dominated by the spectral shift, which is captured by the first adapted feature (AF1). Higher-order AFs progressively encode finer spectral variations and ultimately become dominated by noise and local fluctuations. Principle Component Analysis (PCA), Transfer Component Analysis (TCA), Joint Probability Domain Adaptation (JPDA) [56], and Maximum Independence Domain Adaptation (MIDA) with a linear kernel [57] are benchmarked via root-mean-square error (RMSE) between the AF1 response and PCA-derived generalized spectral shift (both normalized). The latter is calculated from the WGM shifts of all microresonators in the channel, each of which being calculated from spectrally resolved data. Benchmark results for different methods and feature number are summarized in Fig. 3a–f.

With fewer than 20 cavities, all adaptation methods yield RMSE values that can exceed 0.5. Increasing cavities number to 60 reduces the median RMSE to $\approx$0.05 for PCA, TCA, and MIDA, with outliers remaining below 0.15. JPDA performs notably worse, exhibiting a median RMSE of $\approx$0.07, larger wavelength-dependent variability, and higher outlier levels. Its performance does not improve even at the infinite number of features. Among the remaining methods, PCA, MIDA, and TCA exhibit progressively better performance with ultimate RMSE=0.01 with no outliers for TCA. Having 100 microresonators in the channel, TCA achieves a median RMSE of 0.037, a maximum error of $\approx$0.055 (excluding outliers), and the smallest variance across wavelengths. Selecting 5% deviation threshold as acceptable gives wavelength- and cavity-set-independent consistency already with 100 cavities in channel. The RMSE will gradually improve toward 0.01 ("Inf" case) with accumulation of chip responses. Having two different physical channels of the same function, the accuracy will be determined as for 200 microcavities, with RMSE=0.027 and upper deviations <0.04.

The DA procedure aligns the source and target in absolute scale, while agreement with the generalized spectral shift is possible by normalizing the responses. To harmonize signals across chips and channels, AF1 is calibrated after DA. As common in biosensing, the analyte (ANT) dissolved in buffer (PBS) is measured as the sequence of REG-PBS-ANT-PBS phases, where REG denotes the predefined regeneration solution. Owing to the high consistency of the REG–PBS transition, which reflects a bulk refractive index change, the AF1 difference between these two states is used as a calibration reference. Figures 3g,h compare calibrated AF1 and generalized spectral shifts during the ANT–PBS phase for representative measurements (1 $\mu$g/ml IgGH, A-IgGH channels on two different chips). AF1 closely matches the generalized spectral shift in both shape and magnitude, with deviations within confidence interval across different wavelengths. The closely coinciding responses across chips confirm that DA produces robust, chip-independent dynamical signatures. Summing up, the proposed DA makes the selection of the excitation wavelength irrelevant, eliminating the need for spectral scanning and thus boosting temporal resolution from seconds to milliseconds for true real-time detection. This significantly improves the flexibility in selection of affordable excitation sources opening path for distributed accumulation of consistent sensing responses.

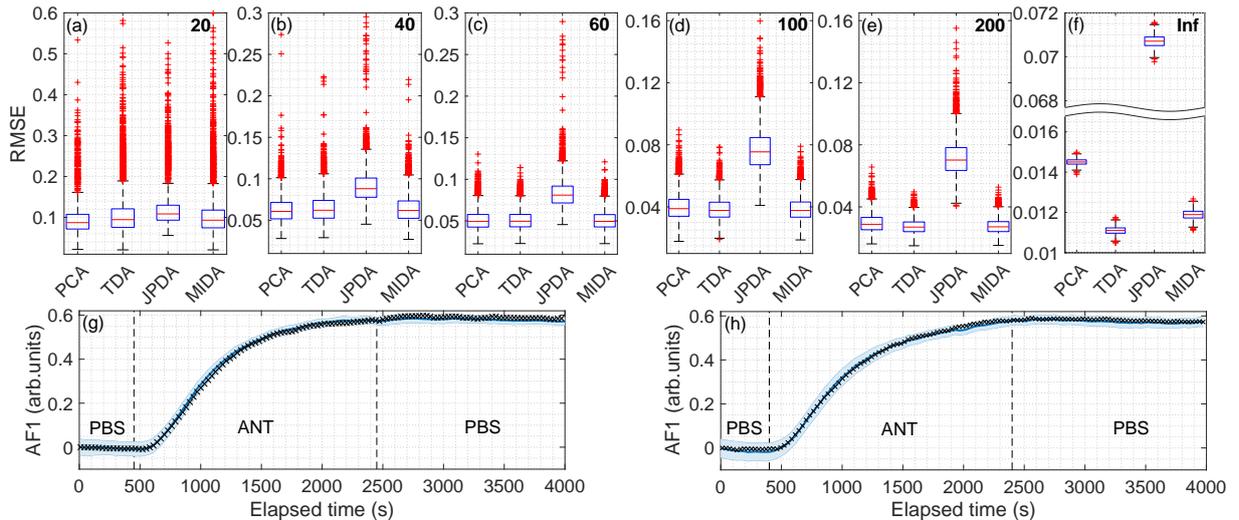

**Fig. 3**: **Sensing domain adaptation.** (a-e) Consistency of the first adapted feature (AF1) with the generalized spectral shift between wavelengths with varied number of cavities (20-200). (f) Results at infinite ("Inf") case when all cavities and wavelengths are merged together as separate features. (g,h) Consistency of the domain adaptation procedure with the spectrally-resolved data processing (black crosses) for two different chips measuring the same biological analyte. Solid blue line shows mean value of the signal in the adapted domain, shaded area - 95% confidence interval.

## Multiplexed detection

Accurate and reproducible biochemical measurements require stabilization or monitoring of key external parameters, including temperature, pH, incubation time, non-specific binding and cross-reactivity. Variations in temperature and pH affect reaction kinetics and binding affinities, while non-specific adsorption (e.g., electrostatic or hydrophobic interactions) can induce false positives and reduce assay specificity. To ensure uniform reaction conditions, incubation time was fixed at 2000 s and freshly prepared PBS was used to minimize fluctuations in pH and ionic strength.

Temperature drift and non-specific adsorption were actively monitored using dedicated sensing channels: thermal sensing was implemented by isolating a channel from the liquid environment (Supplementary Information, Sec. S2), while channel with bare (non-functionalized) microspheres served as reference for non-specific binding.

We used nine different chips (five with 4, two with 9, and two with 16 channels) each including two functional channels (A-IgGR, A-IgGH) and two controls (bare/non-functionalized and isolated/temperature). For 9 or 16 channel chips, same functions were assigned to multiple channels and their responses were merged. A single sensing run is a PBS–ANT–PBS–REG sequence of phases (2000 s per phase), where IgGH and IgGR solutions, individually and in mixtures, served as analytes (ANT). While some chips were employed as disposable devices, three (one per configuration) were restored and re-functionalized. Thus, the full data accounts for variations in chip response, regeneration efficiency, and functionalization consistency.

The responses exhibiting minimal temperature alterations were selected from the entire dataset to avoid manual signal adjustment using the temperature channel (Supplementary Information, Sec. S2). Nonetheless, the entire dataset, including thermally affected measurements, is employed in evaluation of the DL-model performance later. Two functional and one control channel responses are shown for representative examples of different solutions in Fig. 4 a–e. They demonstrate two main phases (ANT–PBS), with a prior PBS phase ($\approx$ 400 s) that describes the transport of solution to the sensing chamber. Functional channels indicate responses to their complementary IgG targets, reflecting high specificity, minimal nonspecific adsorption, and negligible cross-reactivity between IgGH and IgGR. Single-component solutions yield distinct dynamic responses in the corresponding functional channel only (Fig. 4 a,b), whereas mixed analytes activate both functional channels (Fig. 4 c–e). PBS baseline change after analyte binding and its stability indicates durable receptor-analyte linkage. The bare channel remains inactive, overlapping with the non-complementary functional channel in single-analyte and low-concentration mixtures (Fig. 4 a–c). At mixtures with higher concentrations of individual components, it exhibits a step-like response due to bulk refractive index change, which rises with concentration (Fig. 4 d,e). High analyte concentrations may induce weak signals in non-complementary channels due to limited stock solution purity (IgGR $\geq$ 80%, IgGH $\geq$ 95%). The probability for nonspecific adsorption rises with concentration and partial desorption with signal decay upon PBS exchange may be observed (Fig. 4 e).

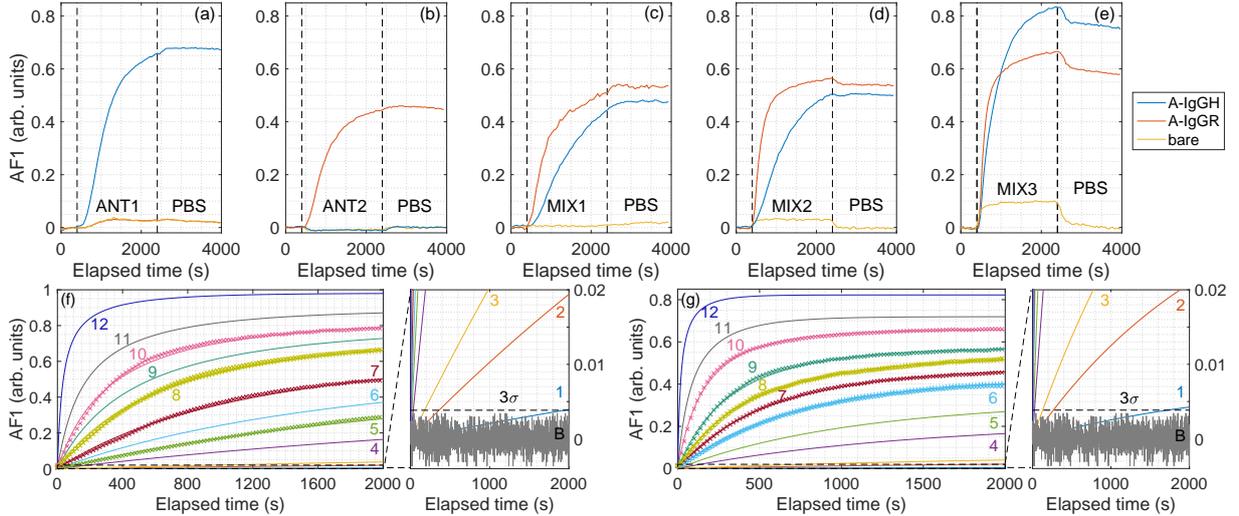

**Fig. 4**: **Multiplexed biosensing with optical microresonator chips.** (a-e) Multichannel chip responses for single- (ANT1: 1 µg/ml IgGH (a), ANT2: 500 ng/ml IgGR (b)) and multi-component (MIX1: 200 ng/ml IgGH + 1 µg/ml IgGR (c), MIX2: 500 ng/ml IgGH + 2 µg/ml IgGR (d), MIX3: 5 µg/ml IgGH + 5 µg/ml IgGR (e)) solutions. (f-g) Functional channel response consistency in the adapted domain among chips for IgGH (f) and IgGR (g) solutions of various concentrations (PBS (B), 1 ng/ml (1), 5 ng/ml (2), 10 ng/ml (3), 50 ng/ml (4), 100 ng/ml (5), 200 ng/ml (6), 500 ng/ml (7), 1 µg/ml (8), 2 µg/ml (9), 5 µg/ml (10), 10 µg/ml (11), 50 µg/ml (12)). Points represent the measurements with different symbols encoding the measurement repetition and lines - fitted model.

We confined the focus to responses of complementary functional channels for single-component solutions to evaluate the sensing performance. Each analyte binding phase is fitted with an adapted adsorption model (Supplementary Information, Sec. S3), enabling determination of kinetic parameters ($k_a$ and $k_d$) and detection limits. Experimental data (points) shows strong agreement with the model (lines), yielding RMSE below $3 \times 10^{-4}$ and a relative variability in parameters across concentrations under 5%. For IgGR, the mean parameter values are $k_a = 2.7 \times 10^5 \ M^{-1}s^{-1}$, $k_d = 3.8 \times 10^{-4} s^{-1}$, and $s_\theta = 1.81$, whereas for IgGH $k_a = 1.9 \times 10^5 \ M^{-1}s^{-1}$, $k_d = 1 \times 10^{-4} s^{-1}$, and $s_\theta = 2.04$ (Fig. 4 f,g). The model remained consistent across the entire concentration range and between measurements using different chips at 0.2 and 1 $\mu$g/ml IgGR and 1 $\mu$g/ml IgGH (various symbols in Fig.4 f,g). This confirms the responses consistency for different chips and acquired at different days. To estimate the detection limit ($LOD$), we reconstructed the responses for missing concentrations within range [1 ng/ml; 50 $\mu$g/ml] and used the approach proposed in [58, 59]. The baseline standard deviation ($\sigma$) was obtained for PBS phase, and the sensitivity ($S$) was derived from a linear fit of signal magnitude at 25 min after start of the binding. Based on $LOD = 3\sigma/S$, the detection limit of the sensor is estimated as ≈6 pM for both IgGR and IgGH.

## Deep-learning processing

Analytical models can describe responses in the adapted domain, but they struggle to incorporate complex, correlated features. For this reason, we considered only single-component and temperature-stable responses while fitting (Fig. 4). Studying mixtures exhibiting cross-linking, accounting for temperature or pH-associated alterations or analyte list expansion would require permanent refinements of the analytical model. Moreover, environmental fluctuations, local perturbations, and missing data introduce noise that substantially degrades analytical accuracy, while the intrinsically nonlinear nature of the sensing responses limits model scalability. Ultimately, this leads to the inability to describe all processes within an analytical model. In contrast, ML architectures naturally accommodate high-dimensional, nonlinear, and noisy data and improve with increasing data volume, making them a more robust, flexible, and scalable solution for multiplexed biosensing.

Exemplary sensing runs affected by thermal instabilities, each comprising ≈4800 technical replicates, are shown in Fig. 5a–c. Temperature fluctuations project coherently across all channels and dominate the responses at low analyte concentrations, particularly in the A-IgGR (Fig. 5b) and A-IgGH (Fig. 5c) channels. For single-analyte solutions, the bare channel closely follows the non-complementary functional channel within the 95% confidence interval (Fig. 5a,b). The bare channel shows no detectable non-specific binding and except for high analyte concentrations, its response reflects preliminary temperature variations and exhibits an inverse correlation with the isolated temperature channel (Supplementary Information, Sec. S2).

Our hybrid DL-framework BioCCF consists of two units: a classifier to identify the solution composition and regressor that estimates concentrations of individual components (Supplementary information, Sec. S4). Classifier filters out analytes that are not present in the solution, then the predicted classes are used together with the channel responses for regressor training. Given the available dataset, only two concentrations (IgGH and IghR) compose the BioCCF output, but other biochemical analytes can be integrated by adding corresponding modules without retraining the entire system. To prevent data leakage and ensure realistic accuracy estimates, we used a custom cross-validation strategy. Each sensing run and all its technical replicates were treated as a set of independent observations. Cross-validation was performed across all sensing runs, with the number of folds matching them. Each fold uses one sensing run for validation and all remaining for training. The learning procedure was executed independently for each sensing run, and the results were subsequently merged, ensuring unbiased benchmarking on previously unseen data.

Complete dataset, acquired from over 200 hours of experimental measurements, comprises 244680 sets of responses including technical replicates. We defined four classes (IgGH, IgGR, MIX and NONE) according to the given dataset, where to reflect the NONE class, the dataset was extended with 96580 chip responses in PBS. Before training BioCCF network, we have excluded first 400 s (pure PBS) from all responses and kept 2000 s of measured responses to improve the DL-engine performance. The performance of BioCCF classification backbone is summarized in Fig. 5 d. The confusion matrix shows well balance in the dataset with ≈80000 responses per class. The classifier achieves an average accuracy of 99.3%, ensuring negligible error propagation to regressor. The NONE class was identified with absolute accuracy and complete separation between single-component solutions was achieved. The lowest accuracy of 97.8% is identified for MIX

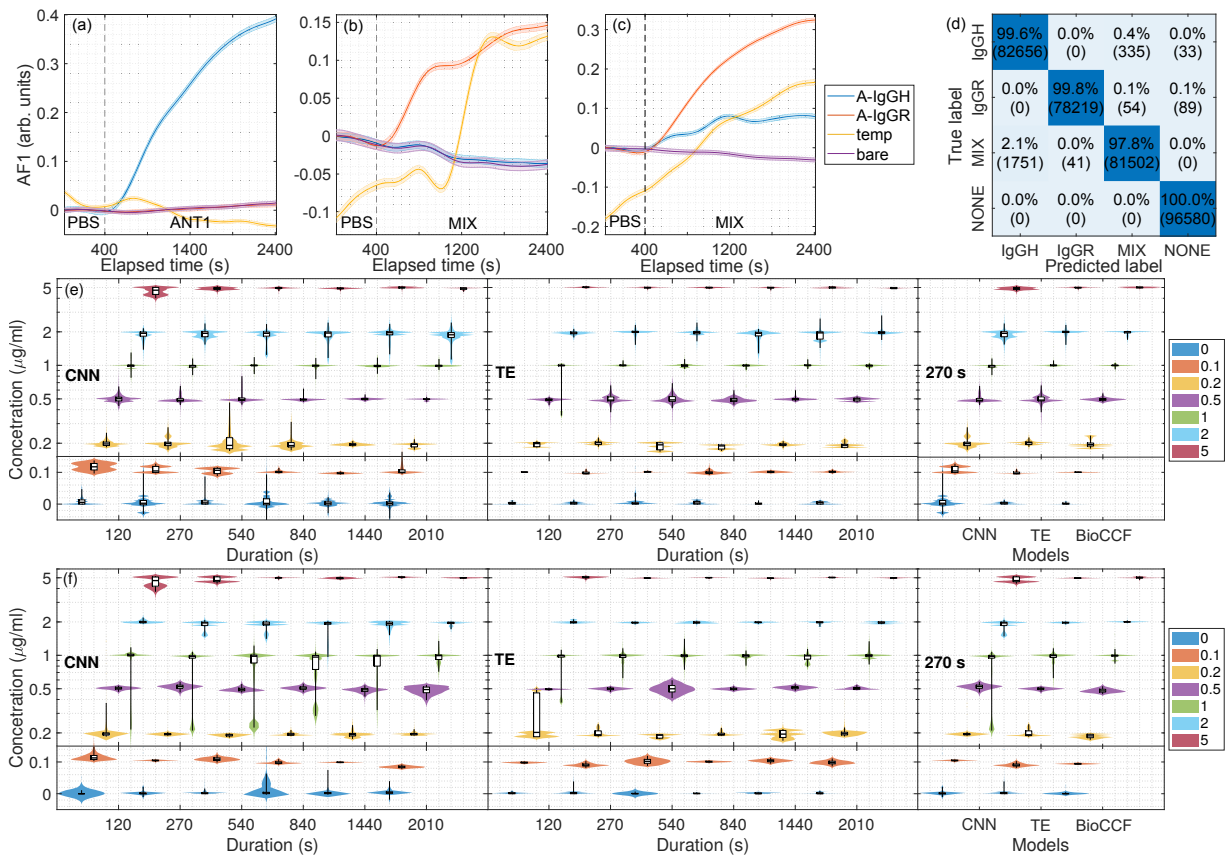

**Fig. 5**: **ML-powered multiplexed biosensing.** (a-c) Representative chip sensing runs for the single phase (ANT1: 1 $\mu$g/ml IgGH (a), ANT2: 0.1 $\mu$g/ml IgGR (b), MIX: 0.1 $\mu$g/ml IgGH + 0.5 $\mu$g/ml IgGR (c)). The response is calculated over $\approx$4800 technical replicates (solid line - mean value, shaded area - 95% confidence interval). (d) Confusion matrix for solution composition classification. (e,f) Statistics on prediction accuracies of the individual concentrations for IgGH (e) and IgGR (f) depending on the response duration used in training. CNN, TE and BioCCF-based ML engines are benchmarked.

samples, which may denote the limited repetitions of all possible mixture combinations.

We benchmarked BioCCF with more common regressor architectures on convolutional neural network (CNN) and Transformer Encoder (TE) (details about architectures in Supplementary information, Sec. S4). While CNN and TE configurations processed channels independently, BioCCF learns interactions between channels that becomes significant with the multiplexing level, particularly for mixed solutions. CNN and TE networks were trained and validated on different durations of sensing responses to determine the shortest one enabling reliable concentration discrimination without overlap. Then, BioCCF has been compared with CNN and TE at optimal response duration.

TE model outperforms the CNN independent on duration. The longer response is available for training, the less is the concentration misclassification rate and the prediction accuracy improves (Fig. 5 e,f). CNN shows serious misclassification errors, where 1 $\mu$g/ml IgGR for up to 1440 s long sequence is predicted as 0.2 $\mu$g/ml and partial misclassification for 2 and 0.1 $\mu$g/ml IgGH is observed for the full sequence (2000 s). TE-based model excludes misclassification starting from the 270 s sequence, thus the latter is defined as sufficient to predict the analyte concentrations (taking into account their limited number) and used for training BioCCF model. The misclaffication for BioCCF is also excluded and the prediction accuracy improves compared to TE. Mean prediction error based upon the median concentration value for IgGH is 35, 8, and 2 ng/ml and for IgGR is 81, 23, and 6 ng/ml for CNN, TE and BioCCF, correspondingly. This means that the cross-channel fusion allows to improve the prediction accuracy by at least 4 times so that for the given measurement range the relative prediction error is $10^{-4}$. The variance for individual concentrations decreases for all values except for IgGH 0.2 $\mu$g/ml and IgGR 0.5 $\mu$g/ml that did not change. The 95% variance range among concentrations improved from 18 to 10 ng/ml for IgGH and from 45 to 14 ng/ml for IgGR solutions, 2 and 4 times, correspondingly. For the baseline and the smallest concentration, the 95% variance range does not exceed 4 ng/ml. Thus, BioCCF effectively handles correlated sensing responses and environmental fluctuations in multiplexed biosensing. Its performance is restricted by the detection limit only, which was quantified using the analytical model fitting to the adapted single-component responses under strictly controlled experimental conditions.

## Discussion

The combination of the affordable, modular WGM glass chip integrating tens of thousands of high-Q microresonators across up to 100 parallel sensing channels with a hybrid ML-engine establishes a new paradigm for high-throughput multiplexed optical biosensing. Domain-adapted ML-framework enables chip-independent, rapid, and accurate concentrations estimations for individual analytes in complex solutions, eliminating the need for chip-specific recalibration. This approach allows consistent accumulation of sensing data across heterogeneous devices, overcoming fundamental limitations imposed by variability in manufacturing high-sensitivity devices. Simplicity and affordability of the chip fabrication support disposable sensing formats, mitigating other barriers to practical deployment - the limited stability and reusability of biorecognition layers.

Immunoassay experiments demonstrate prediction of IgG components in complex mixtures within <5 min and with relative prediction error of $10^{-4}$ for up to ∼10 $\mu$g/ml concentrations, validated on nine chips over numerous sensing runs. Importantly, this does not represent the ultimate performance capability of the sensor. Optimization of the biorecognition layers, where random immobilization is replaced with controlled receptor orientation, is expected to improve detection limits by orders of magnitude. Alternative strategy is to use artificial antibodies (aptamers) as recognition layers. They have longer lifetime and are robuster to environmental variations, but due to resource-intensive and time-consuming sequence selection process are limitedly accessible to the sensing community. By accumulating large, consistent, and independently measured datasets, our sensing platform provides a powerful pathway to accelerate, benchmark, and validate emerging bioreceptors.

Beyond glass microspheres used as elementary sensing units in this study, the platform supports integration of microresonators of different geometries and materials within a single chip. Examples include hybrid integration of glass and PMMA microspheres [20] together with synthesised polymer optical microtodoids [17, 19, 60], enabling extended functionality and multimodal sensing. These microresonators will be treated as additional sensing channels, naturally supported by the scalability of the cross-channel fusion implemented in the ML-engine. The latter is transferable to other optical biosensing technologies that generate sensorgrams, such as recently demonstrated massively multiplexed surface plasmon resonance (SPR) [61]. In this context, the ML-framework has the potential to serve as a universal processing backbone for optical biosensors once data consistency is ensured.

The novel vision of the massive sensor multiplexing opens a pathway for rapid technology expansion and distributed, large-scale data acquisition across the community. Shared repositories containing diverse sensing datasets, spanning multiple analytes, biorecognition layers, experimental conditions, and laboratories, can continuously enhance the accuracy, robustness, and generalizability of the DL framework. Such an ecosystem will enable direct sharing of raw biosensing data and establish the foundation for a scalable, DL-enabled paradigm in optical biosensing.

# Materials and methods

## Chemicals

Phosphate-buffered saline (PBS) pH = 7.4 was prepared in deionized water containing 0.01 M phosphate buffer, 0.0027 M potassium chloride, and 0.137 M sodium chloride. Two immunoglobulin G (IgG) species — rabbit IgG (I8140) and human serum IgG (I2511) — were selected as target analytes (ANT), while rabbit (F9887) and human (F5512) anti-IgG (A-IgG) molecules served as bioreceptors enabling selective antibody–antigen binding. Surface functionalization of microcavities employed APTMS ((3-aminopropyl)trimethoxysilane, 281778) to introduce amino ($-NH_2$) groups, followed by antibody immobilization via glutaraldehyde (GA, G7526) cross-linking. A 1 mM NaOH (S5881) solution was used for sensor regeneration (REG), enabling multiple measurement phases within a single experimental run. All chemicals were obtained from Sigma-Aldrich. Bovine serum albumin (BSA, Thermo Scientific 37525) was diluted to 0.01% and used as a blocking agent to minimize nonspecific adsorption.

## WGM microresonators

Soda-lime glass microsphere (Cospheric LLC) is the elementary sensing unit of the multiplexed chip. Unlike pure silica, soda-lime glass exhibits better chemical and thermal stability, attributed to the presence of sodium carbonate and calcium carbonate. The majority of microspheres in the supplied batch range in size from 106 to 125 $\mu$m. Before chip fabrication, the microspheres undergo cleaning that involves ultrasonication in methanol for 20 minutes, centrifugation to remove surface contaminants and drying.

## Chip functionalization

The functionalization process begins with chips cleaning to remove dust and residues remaining after fabrication and then treating with low-pressure oxygen plasma to eliminate organic contaminants and generate surface hydroxyl (–OH) groups. Immediately after activation, vapor phase silanization is performed, which ensures uniform coating and precise control of surface modification compared to the wet-phase process. The chips are subsequently rinsed with 2-propanol and ultrapure water to remove unbound silane molecules and dried under nitrogen flow. The silanized chips are processed with 2.5% GA in PBS for 30 min at room temperature and then rinsed with ultrapure water. Channel specificity is ensured by spotting droplets of bioreceptors (A-IgG) solutions and incubation for 1 h at room temperature to allow immobilization. Finally, the chips are rinsed with PBS to remove unbound receptors and treated with a BSA blocking solution.

## Instrument

WGM signals are excited via an optical prism and radiative emission of microresonators is monitored. A monochrome high-speed global shutter camera (CB262RG-GP-X8G3, Ximea) enables spatially resolved acquisition of signals from individual microresonators. Uniform excitation across the chip is ensured by a wide-field laser beam ($\approx$ 8 mm diameter) collimated with achromatic fiber package (60FC-T-4-M40-24, Schäfter+Kirchhoff). Beam elongation at the prism interface was compensated by transforming the beam profile from circular to elliptical. With this excitation and detection scheme plenty of microcavities are excited and tracked at once. The excitation source is a tunable diode laser (Velocity, New Focus; 680–690 nm, 200 kHz linewidth), coupled into a single-mode fiber. Light polarization is controlled to excite TE modes, and the wavelength is monitored by a wavemeter (WS7-30, HighFinesse).

WGM chip is mounted onto the prism via immersion oil and assembled with microfluidic flow chamber to construct a sensing head. The reusable chamber, fabricated from polyester ketone ester, contains a single inlet–outlet configuration forming one continuous flow path covering all resonators. A transparent window is incorporated to allow real-time monitoring of fluid motion and to enable WGM signal collection. Analyte delivery is ensured by a pressure-driven controller (LINEUP FLOW EZ, Fluigent) combined with a flow-rate sensor (FLOW UNIT, Fluigent) and multiport valve (M-SWITCH, Fluigent) to enable sequential injection of solutions.

# Acknowledgements

The authors Andreas Ostendorf and Anton Saetchnikov are grateful to the German Federal Ministry of Research, Technology and Space (BMFTR) for partly funding this work under the VIP+-Programme in the project IntellOSS, 03VP08220.

# Contributions

A.S conceived the idea and designed the project. I.S. and A.S. developed theoretical framework and performed data analysis. A.S. fabricated multiplexed chips and carried out the experiments. I.S., E.T. and A.S. developed data processing methodology. I.S. designed the neural network architecture and implemented the deep-learning engine. I.S., E.T., A.O., and A.S. contributed to the discussions, theoretical analysis, data validation and interpretation. A.O. provided resources for experimental investigations. A.S. supervised and coordinated the overall research. I.S. and A.S. wrote original draft. All authors discussed the results and contributed to reviewing and editing the manuscript.

# Data availability

The data that support the findings of this study are available from the corresponding author on reasonable request.

# Conflict of interest

The authors declare no conflicts of interest.

# References

[1] Braginsky, V. B., Gorodetsky, M. L. & Ilchenko, V. S. Quality-factor and nonlinear properties of optical whispering-gallery modes. *Physics Letters A* **137**, 393–397 (1989).

[2] Vahala, K. J. Optical microcavities. *Nature* **424**, 839–846 (2003).

[3] Jiang, X., Qavi, A. J., Huang, S. H. & Yang, L. Whispering-gallery sensors. *Matter* **3**, 371–392 (2020).

[4] Yu, D. *et al.* Whispering-gallery-mode sensors for biological and physical sensing. *Nature Reviews Methods Primers* **1**, 453 (2021).

[5] Vollmer, F. & Arnold, S. Whispering-gallery-mode biosensing: Label-free detection down to single molecules. *Nature methods* **5**, 591–596 (2008).

[6] Zhu, J. *et al.* On-chip single nanoparticle detection and sizing by mode splitting in an ultrahigh-q microresonator. *Nature Photonics* **4**, 46–49 (2010).

[7] Shao, L. *et al.* Detection of single nanoparticles and lentiviruses using microcavity resonance broadening. *Advanced materials (Deerfield Beach, Fla.)* **25**, 5616–5620 (2013).

[8] He, L., Ozdemir, S. K., Zhu, J., Kim, W. & Yang, L. Detecting single viruses and nanoparticles using whispering gallery microlasers. *Nature nanotechnology* **6**, 428–432 (2011).

[9] Baaske, M. D., Foreman, M. R. & Vollmer, F. Single-molecule nucleic acid interactions monitored on a label-free microcavity biosensor platform. *Nature nanotechnology* **9**, 933–939 (2014).

[10] Chen, W., Kaya Özdemir, Ş., Zhao, G., Wiersig, J. & Yang, L. Exceptional points enhance sensing in an optical microcavity. *Nature* **548**, 192–196 (2017).

[11] Reynolds, T. *et al.* Fluorescent and lasing whispering gallery mode microresonators for sensing applications. *Laser & Photonics Reviews* **11**, 1600265 (2017).

[12] Toropov, N. *et al.* Review of biosensing with whispering-gallery mode lasers. *Light: Science & Applications* **10**, 42 (2021).

[13] Chen, Y., Yin, Y., Ma, L. & Schmidt, O. G. Recent progress on optoplasmonic whispering–gallery–mode microcavities. *Advanced Optical Materials* **9**, 2100143 (2021).

[14] Houghton, M. C., Kashanian, S. V., Derrien, T. L., Masuda, K. & Vollmer, F. Whispering-gallery mode optoplasmonic microcavities: From advanced single-molecule sensors and microlasers to applications in synthetic biology. *ACS Photonics* **86**, 37 (2024).

[15] Liao, J. & Yang, L. Optical whispering-gallery mode barcodes for high-precision and wide-range temperature measurements. *Light: Science & Applications* **10**, 32 (2021).

[16] Stoian, R.-I., Lavine, B. K. & Rosenberger, A. T. pH sensing using whispering gallery modes of a silica hollow bottle resonator. *Talanta* **194**, 585–590 (2019).

[17] Saetchnikov, A. V., Tcherniavskaia, E. A., Saetchnikov, V. A. & Ostendorf, A. Detection of per- and polyfluoroalkyl water contaminants with a multiplexed 4D microcavities sensor. *Photonics Research* **11**, A88 (2023).

[18] Tang, S.-J. *et al.* Single-particle photoacoustic vibrational spectroscopy using optical microresonators. *Nature Photonics* **17**, 951–956 (2023).

[19] Saetchnikov, A. V., Tcherniavskaia, E. A., Saetchnikov, V. A. & Ostendorf, A. Two-photon polymerization of optical microresonators for precise ph sensing. *Light: Advanced Manufacturing* **5**, 1 (2024).

[20] Saetchnikov, I., Tcherniavskaia, E., Ostendorf, A. & Saetchnikov, A. Induced eccentricity splitting in disordered optical microspheres for machine learning enabled wavemeter. *Arxiv* **2412** (2024).

[21] Suebka, S., Gin, A. & Su, J. Frequency locked whispering evanescent resonator (flower) for biochemical sensing applications. *Nature protocols* **20**, 1616–1650 (2025).

[22] Gather, M. C. & Yun, S. H. Single-cell biological lasers. *Nature Photonics* **5**, 406–410 (2011).

[23] Humar, M. & Yun, S. H. Intracellular microlasers. *Nature Photonics* **9**, 572–576 (2015).

[24] Martino, N. *et al.* Wavelength-encoded laser particles for massively multiplexed cell tagging. *Nature Photonics* **13**, 720–727 (2019).

[25] Kavčič, A. *et al.* Deep tissue localization and sensing using optical microcavity probes. *Nature communications* **13**, 1269 (2022).

[26] Loyez, M., Adolphson, M., Liao, J. & Yang, L. From whispering gallery mode resonators to biochemical sensors. *ACS sensors* **8**, 2440–2470 (2023).

[27] Hao, S. & Su, J. Whispering gallery mode optical resonators for biological and chemical detection: Current practices, future perspectives, and challenges. *Reports on progress in physics. Physical Society* **88** (2024).

[28] Baker, M. Reproducibility crisis: Blame it on the antibodies. *Nature* **521**, 274–276 (2015).

[29] Vollmer, F., Arnold, S., Braun, D., Teraoka, I. & Libchaber, A. Multiplexed dna quantification by spectroscopic shift of two microsphere cavities. *Biophysical Journal* **85**, 1974–1979 (2003).

[30] Mallik, A. K. *et al.* Whispering gallery mode micro resonators for multi-parameter sensing applications. *Optics express* **26**, 31829–31838 (2018).

[31] Yan, Y.-Z. *et al.* Robust spot-packaged microsphere-taper coupling structure for in-line optical sensors. *IEEE Photonics Technology Letters* **23**, 1736–1738 (2011).

[32] Niu, P. *et al.* Hollow-microsphere-integrated optofluidic immunochip for myocardial infarction biomarker microanalysis. *Biosensors & bioelectronics* **248**, 115970 (2024).

[33] Kavungal, V. *et al.* Packaged inline cascaded optical micro-resonators for multi- parameter sensing. *Optical Fiber Technology* **50**, 50–54 (2019).

[34] Monifi, F., Ozdemir, S. K., Friedlein, J. & Yang, L. Encapsulation of a fiber taper coupled microtoroid resonator in a polymer

matrix. *IEEE Photonics Technology Letters* **25**, 1458–1461 (2013).

[35] Gin, A. *et al.* Label-free, real-time monitoring of membrane binding events at zeptomolar concentrations using frequency-locked optical microresonators. *Nature communications* **15**, 7445 (2024).

[36] Schumann, M., Bückmann, T., Gruhler, N., Wegener, M. & Pernice, W. Hybrid 2d–3d optical devices for integrated optics by direct laser writing. *Light: Science & Applications* **3**, e175 (2014).

[37] Kelemen, L. *et al.* Direct writing of optical microresonators in a lab-on-a-chip for label-free biosensing. *Lab on a chip* **19**, 1985–1990 (2019).

[38] Ouyang, X. *et al.* Ultrasensitive optofluidic enzyme-linked immunosorbent assay by on-chip integrated polymer whispering-gallery-mode microlaser sensors. *Lab on a chip* **20**, 2438–2446 (2020).

[39] Bog, U. *et al.* On-chip microlasers for biomolecular detection via highly localized deposition of a multifunctional phospholipid ink. *Lab on a chip* **13**, 2701–2707 (2013).

[40] Álvarez Freile, J., Choukrani, G., Zimmermann, K., Bremer, E. & Dähne, L. Whispering gallery modes-based biosensors for real-time monitoring and binding characterization of antibody-based cancer immunotherapeutics. *Sensors and Actuators B: Chemical* **346**, 130512 (2021).

[41] Suebka, S., McLeod, E. & Su, J. Ultra-high-q free-space coupling to microtoroid resonators. *Light: Science & Applications* **13**, 75 (2024).

[42] de Vos, K., Bartolozzi, I., Schacht, E., Bienstman, P. & Baets, R. Silicon-on-insulator microring resonator for sensitive and label-free biosensing. *Optics express* **15**, 7610–7615 (2007).

[43] Claes, T. *et al.* Label-free biosensing with a slot-waveguide-based ring resonator in silicon on insulator. *IEEE Photonics Journal* **1**, 197–204 (2009).

[44] Cognetti, J. S. *et al.* Disposable photonics for cost-effective clinical bioassays: Application to covid-19 antibody testing. *Lab on a chip* **21**, 2913–2921 (2021).

[45] Ji, X., Roberts, S., Corato-Zanarella, M. & Lipson, M. Methods to achieve ultra-high quality factor silicon nitride resonators. *APL Photonics* **6**, 12987 (2021).

[46] Barrios, C. A. Integrated microring resonator sensor arrays for labs-on-chips. *Analytical and bioanalytical chemistry* **403**, 1467–1475 (2012).

[47] Tcherniavskaia, E. A. & Saetchnikov, V. A. Application of neural networks for classification of biological compounds from the characteristics of whispering-gallery-mode optical resonance. *Journal of Applied Spectroscopy* **78**, 457–460 (2011).

[48] Lu, J. *et al.* Experimental demonstration of multimode microresonator sensing by machine learning. *IEEE Sensors Journal* **21**, 9046–9053 (2021).

[49] Li, Z. *et al.* Smart ring resonator–based sensor for multicomponent chemical analysis via machine learning. *Photonics Research* **9**, B38 (2021).

[50] Duan, B. *et al.* High-precision whispering gallery microsensors with ergodic spectra empowered by machine learning. *Photonics Research* **10**, 2343 (2022).

[51] Chen, Q. *et al.* Optical frequency comb-based aerostatic micro pressure sensor aided by machine learning. *IEEE Sensors Journal* **23**, 21078–21083 (2023).

[52] Saetchnikov, A., Tcherniavskaia, E., Saetchnikov, V. & Ostendorf, A. Deep-learning powered whispering gallery mode sensor based on multiplexed imaging at fixed frequency. *Opto-Electronic Advances* **3**, 200048

(2020).

[53] Saetchnikov, A. V., Tcherniavskaia, E. A., Saetchnikov, V. A. & Ostendorf, A. Intelligent optical microresonator imaging sensor for early stage classification of dynamical variations. *Advanced Photonics Research* **7**, 2100242 (2021).

[54] Zossimova, E. *et al.* Whispering gallery mode sensing through the lens of quantum optics, artificial intelligence, and nanoscale catalysis. *Applied Physics Letters* **125**, 13832 (2024).

[55] Gorodetsky, M. L. & Ilchenko, V. S. Optical microsphere resonators: Optimal coupling to high-Q whispering-gallery modes. *Journal of the Optical Society of America B* **16**, 147 (1999).

[56] Zhang, W. & Wu, D. Discriminative joint probability maximum mean discrepancy (djp-mmd) for domain adaptation (2020). arXiv:1912.00320.

[57] Yan, K., Kou, L. & Zhang, D. Learning domain-invariant subspace using domain features and independence maximization. *IEEE transactions on cybernetics* **48**, 288–299 (2018).

[58] Harris, D. C. *Quantitative chemical analysis* 6th ed. edn (W. H. Freeman, New York, 2003).

[59] Loock, H.-P. & Wentzell, P. D. Detection limits of chemical sensors: Applications and misapplications. *Sensors and Actuators B: Chemical* **173**, 157–163 (2012).

[60] Saetchnikov, A. V., Tcherniavskaia, E. A., Saetchnikov, V. A. & Ostendorf, A. A laser written 4D optical microcavity for advanced biochemical sensing in aqueous environment. *Journal of Lightwave Technology* **38**, 2530–2538 (2020).

[61] Agu, C. V. *et al.* Multiplexed proteomic biosensor platform for label-free real-time simultaneous kinetic screening of thousands of protein interactions. *Communications biology* **8**, 468 (2025).

[62] Saetchnikov, A., Tcherniavskaia, E., Saetchnikov, V. & Ostendorf, A. *Mapping of the detecting units of the resonator-based multiplexed sensor*, Vol. 10678 of *SPIE Proceedings*, 106780W (2018).

[63] Duda, R. O. & Hart, P. E. Use of the hough transformation to detect lines and curves in pictures. *Commun. ACM* **15**, 11–15 (1972).

[64] Dabov, K., Foi, A., Katkovnik, V. & Egiazarian, K. Bm3d image denoising with shape-adaptive principal component analysis. *Proc. Workshop on Signal Processing with Adaptive Sparse Structured Representations (SPARS'09)* (2009).

[65] Saetchnikov, A. *et al. Ai-based solution for robust detection with optical microresonators*, Vol. 12618 of *SPIE Proceedings*, 1261803 (2023).

[66] Ikotun, A. M., Ezugwu, A. E., Abualigah, L., Abuhaija, B. & Heming, J. K-means clustering algorithms: A comprehensive review, variants analysis, and advances in the era of big data. *Information Sciences* **622**, 178–210 (2023).

[67] Dong, C.-H. *et al.* Fabrication of high-q polydimethylsiloxane optical microspheres for thermal sensing. *Applied Physics Letters* **94**, 839 (2009).

[68] Li, H. & Fan, X. Characterization of sensing capability of optofluidic ring resonator biosensors. *Applied Physics Letters* **97**, 011105 (2010).

[69] Li, H. *et al.* Packaged wgm mbr sensor for high-performance temperature measurement using cnn-based multimode barcode images. *Optics express* **32**, 5515 (2024).

[70] Schaaf, P. & Talbot, J. Surface exclusion effects in adsorption processes. *The Journal of Chemical Physics* **91**, 4401–4409 (1989).

[71] Wilson, K. A., Finch, C. A., Anderson, P., Vollmer, F. & Hickman, J. J. Whispering gallery mode biosensor quantification of

fibronectin adsorption kinetics onto alkylsilane monolayers and interpretation of resultant cellular response. *Biomaterials* **33**, 225–236 (2012).

# Supplementary Information for:

# Rethinking massive multiplexing in whispering gallery mode biosensing

Ivan Saetchnikov[1], Elina Tcherniavskaia[2], Andreas Ostendorf[3], Anton Saetchnikov[3]

[1]Radio Physics Department, Belarusian State University, Minsk, 220064, Belarus

[2]Physics Department, Belarusian State University, Minsk, 220030, Belarus

[3]Chair of Applied Laser Technologies, Ruhr University Bochum, Bochum, 44801, Germany

Corresponding author(s): Anton Saetchnikov. E-mail(s): anton.saetchnikov@rub.de

## S1  Microresonator mapping

Due to the random nature for allocation of the microresonators on the chip their individual positions are required to be identified in order to record accurate sensing data. We interpret this as a computer vision problem in which a large number of similarly shaped objects are detected in a 2D image [62]. The approach is based on a pipeline consisting of various preprocessing steps followed by a circle Hough Transform [63] (Fig. S1). Preprocessing steps include image denoising using the BM3D method [64], image sharpening to improve edge visibility, background correction and histogram equalization (Fig. S1 a-d). The preprocessing outcome is a clear image with high contrast between the background and the microresonators and can be directly used as input for circle Hough Transform. A representative chip with corresponding sensor map is shown in Fig. S1 e. The absolute majority of microresonators are identified, only a few remained non- or incorrectly localized. The identification run averaged over >20 chips with optimized parameters delivers the detection accuracy of >97% with minimum of 95% where errors refer to Hough Transform's tendency to underestimate (Fig. S1 f). Among the detected regions the median number of the falsely identified microreresonators (false detection rate, FDR) remains below 3% (Fig. S1 g). The latter are often associated with the microresonators defects, light refractions and obstructive dust particles [65]. Backlighting variations may cause the accuracy of the finely tuned implementation to drop by more than 50%. Therefore, to avoid the need for regular adjustment of the model parameters, consistent backlighting conditions are maintained.

The groups of microresonators acting as sensing channels form a regular $n \times n$ grid on the chip, determined by the geometry of the pass-through mask used for microcavity deposition. Since the number of channels is predefined for each chip and the microresonator positions are known from the previous step, the channel assignment is realized using k-means clustering [66]. The regularity of the cluster locations enables initialization of the cluster centers by uniformly partitioning the chip area into $n \times n$ sections. These predefined cluster counts and their initial centers serve as the input parameters for the k-means procedure. K-means approach is particularly well suited for this task because it minimizes intra-cluster variance and reliably converges to spatially connected clusters, even when initial estimates are approximate. Moreover, its computational efficiency makes it attractive for high-throughput chip characterization. The algorithm exhibits robust performance on chips with accurately positioned microcavities within the clusters. Outlying microresonators that may be accidentally fixed between the main cluster regions are automatically assigned to the nearest centroid. At the same time, to ensure consistency in the biochemical sensing responses, such outlying cavities are excluded from further analysis.

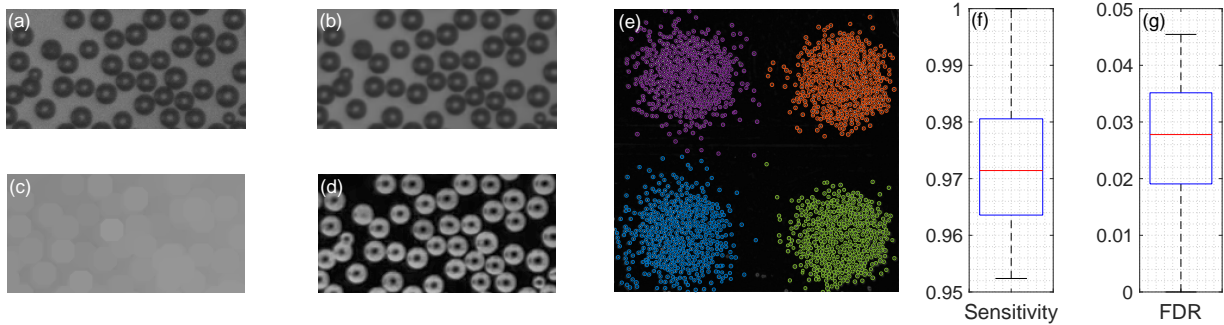

**Fig. S1**: **Microresonator mapping.** (a-d) Multistep preprocessing of the microsphere-based multiplexed chip image. (a) Original image. (b) Denoised image. (c) Estimated background illumination. (d) Final preprocessed image. (e) Overview of identified areas of individual microspheres with color-coded cluster assignment. (f-g) Statistics of the microresonator localization accuracy estimated for >20 multiplexed microresonator chip samples in both sensitivity (f) and FDR (g).

## S2  Temperature control channel

Temperature-induced alterations in the resonator and environmental refractive index (thermo-optical coefficient) as well as resonator physical dimensions (thermal expansion coefficient) disturb the response monitored in single [67, 68] or multiple [15, 69] WGMs. Owing to high coefficient of thermal expansion and negative thermo-optic coefficient, low attenuation loss, good mechanical and chemical stability, polydimethylsiloxane (PDMS) is used for microcavity fabrication [67] or for thin film resonator coating [68] for high-precision thermal sensing. We propose the encapsulation of one of the sensor channels in PDMS by spin-coating to uniformly isolate the cavities from the environment and provide on-chip temperature monitoring. Thus, the mechanism of the thermal response is limited by thermo-optic PDMS changes. PDMS forms a protective hydrophobic layer minimizing interaction with water and reducing non-specific binding of biomolecules. Being transparent in the visible range, it does not hinder signal detection in multiplexed imaging. As long as all microresonators in the channel remain entirely isolated, the remaining inhomogeneity associated with their random distribution does not affect the performance.

The performance of the on-chip temperature channel was evaluated using four sequential PBS–ANT–PBS–REG cycles (2000 s each sensing phase) with 0.1 $\mu$g/ml IgGH as the analyte. External temperature sensor (PT 100, resolution 0.01 K) integrated with the sensing head verifies the spectral response consistency of the PDMS-isolated channel with actual temperature fluctuations. The unscaled response of the isolated channel over almost 10 hours, together with PT100 readings, is shown in Fig. S2 a and reveals a clear inverse correlation between the channel response and actual temperature. Only the first phase shows an anomalous positive trend, attributed to thermal balancing between the fluid and sensing head after the assembly. Combining responses from all biochemical phases together confirms a consistent linear dependence between the isolated-channel signal and temperature (Fig. S2 b), as expected for small thermal variations. The residuals around the linear fit primarily reflects PT100 accuracy limits and hysteresis. Since the confidence interval remains nearly constant along the fit, the biochemical medium has negligible influence on the temperature-channel response. Thus, this method reliably captures thermal fluctuations and can be used to correct biochemical signals for temperature-induced drift.

An example of a complementary functional-channel response for a PBS–ANT–PBS sequence, before (dashed) and after (solid) temperature correction, is shown in Fig. S2 c. Temperature fluctuations with a total span of 0.8 °C noticeably distort the binding dynamics of a relatively high IgGR concentration (2 $\mu$g/ml). This distortion is most evident in the steady-state PBS phases before and after molecular attachment, where correcting for the isolated temperature-channel signal restores temporal consistency. The binding curve itself also becomes more accurate, particularly near saturation, where temperature drift is most pronounced. Overall, incorporating on-chip thermal compensation substantially improves the reliability of functional-channel analysis and enhances sensor performance.

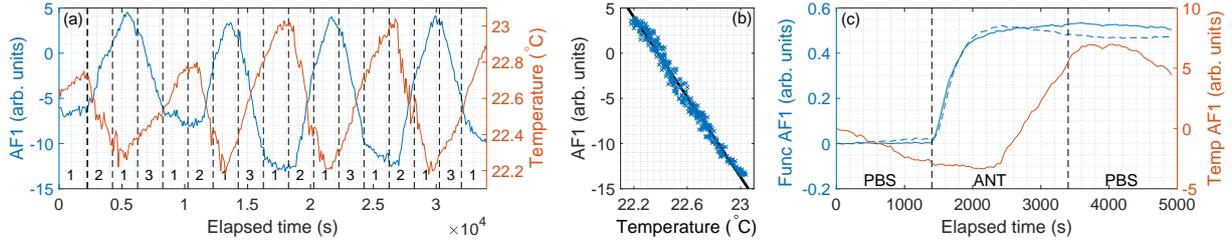

**Fig. S2**: **Temperature control channel.** (a) Spectral response of the PDMS isolated control channel (left axis) along with the temperature variations measured by the external sensor (right axis) in varying biochemical environment (PBS (1), ANT (0.1 µg/ml IgGH) (2) and REG (3)). (b) Linear fit of the channel response to the temperature. (c) The impact of the temperature/isolated channel response (right) for detection of 2 µg/ml IgGR with A-IgGR channel (left) as raw (dashed) and corrected (solid) response curves.

## S3 Small molecule adsoprtion model

We used the modification of the small molecule adsorption model that incorporates the surface exclusion effect indicating the capacity of previously adsorbed particles to hinder the adsorption of subsequent molecules [70, 71]. This model is described analytically as follows:

$$\frac{d\theta}{dt} = k_a C \phi(\theta) - k_d \theta, \tag{S1}$$

where $\theta$ is the surface coverage, $k_a$, $k_d$ are the association and dissociation constants and $C$ is the concentration given in mol/L (M) values. Surface exclusion effects are considered by describing the amount of free binding sites via function $\phi(\theta)$ approximated by the empirical expression:

$$\phi(\theta) = \frac{(1-x)^3}{1 - 0.8120x + 0.2236x^2 + 0.0845x^3}, \tag{S2}$$

where $x = \theta/\theta_\infty$ and $\theta_\infty$ is equal to 0.547 for spherical particles [70].

To operate with sensing data from the adapted domain, denoted as AF, the scaling parameter has been added to the model. It adjusts the surface coverage $\theta_s = s_\theta \theta$ to sensor response.

## S4 BioCCF network model

Novel hybrid ML-framework that we named biological cross-channel fusion (BioCCF) addresses the task of multiplexed analysis of complex biological solutions by combining a classifier for analyte identification with regressor to estimate concentrations of individual analytes (Fig. S3).

The classification backbone employs a hybrid architecture that integrates convolutional neural network (CNN) layers with bidirectional long short-term memory (BiLSTM) layers. The model includes two sequential 1D convolutional modules with 16 and 32 filters, respectively, and kernel sizes decreasing from 5 to 3 while maintaining consistent padding. Each convolutional block is followed by LeakyReLU activation, batch normalization, and dropout layer with probability of 0.3 to mitigate overfitting. The resulting feature representations are processed with BiLSTM layer out of 32 hidden units, followed by another dropout layer (rate = 0.3) and a final output layer for classification. This design enables the model to first capture local spatial dependencies and suppress noise through convolutional operations followed by learning complex temporal relationships in both forward and backward directions by the BiLSTM component.

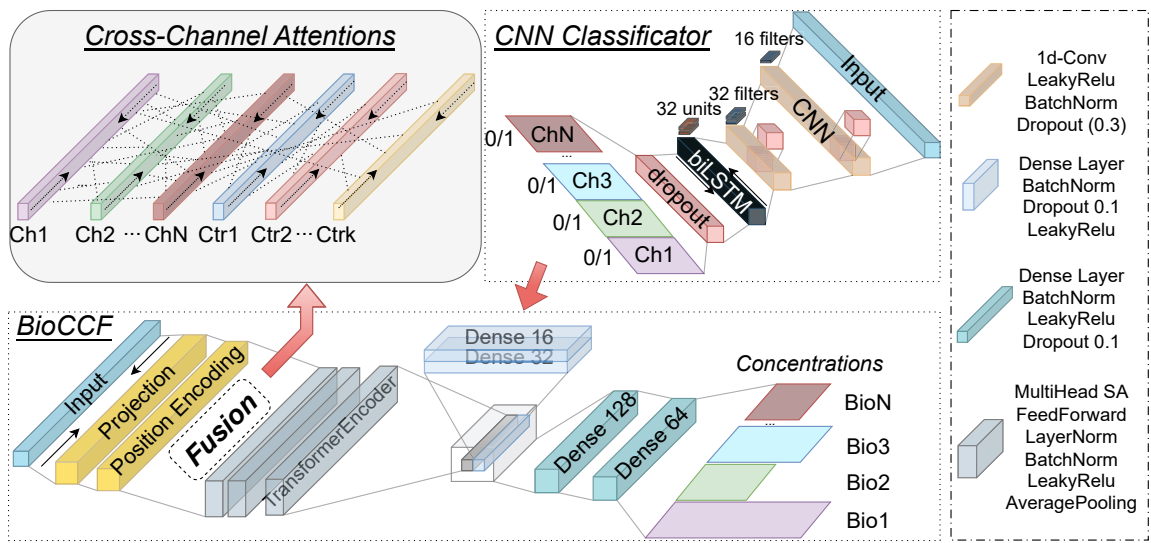

**Fig. S3**: Architecture of the proposed BioCCF network for multiplexed biochemical solutions analysis.

The regression backbone is based on Transformer Encoder model where the channel responses are first passed through the projection block and position encoding block combined with cross-feature fusion attention before applying a set of transformer encoders. Projection layer performs a dense transformation of the input into a 16-dimensional space and position encoding generates unique 16-dimensional vector representations for each temporal position in the input. The outputs from the projection layer and position encodings are then passed through the cross-channel fusion. This block is designed to preprocess and learn interactions between different channels. It applies cross-attention to time sections of the channel responses both within single channel and among different channels using a multi-head attention block with 2 heads. The fused outputs then pass through three sequential transformer encoder blocks, each comprising a 4-head attention layer, a feed-forward network with GELU and dropout activations, and a normalization layer. Between the transformer encoder blocks, batch normalization and LeakyReLU layers are applied to introduce normalization and nonlinearity, alongside average pooling with a kernel size of 2 to reduce sequence lengths. Categorical outputs of the classification backbone are passed through two dense layers of sizes 16 and 32, respectively, each followed by BatchNorm, LeakyReLU, and dropout (0.1). Finally, features extracted from the channel sensorgrams and categorical features are concatenated and passed through two fully connected layers with BatchNorm, LeakyReLU, and dropout, reducing the dimensionality from 128 to 64.

The performance of the novel BioCCF framework for analysing multiplexed biosensing data processing was benchmarked by comparing it with a CNN-based regressor and a Transformer Encoder without integrated cross-channel attention block. The latter except for the absence of the cross-channel attention block has twice reduced the number of heads in the transformer encoder blocks. CNN-based regressor consists of three CNN modules with 32, 64, and 128 filters (kernel sizes of 5, 5, and 3), each followed by BatchNorm, LeakyReLU, and MaxPooling layers, and global average pooling. The extracted features are concatenated with categorical features and processed further according to the BioCCF pipeline. All deep learning models were trained and validated on a hardware with a 14-core Intel Xeon W-2275 processor and 128 GB memory, utilizing NVIDIA Quadro RTX 6000 (24 GB VRAM) GPU to accelerate computation. The development environment was based on Python, employing CUDA for GPU acceleration and leveraging PyTorch as the primary deep learning framework. Data preprocessing, manipulation, and visualization were performed with MATLAB.